\documentclass[11pt]{aastex}
\usepackage{emulateapj5,apjfonts}
\usepackage{onecolfloat}
\usepackage{epsf}
\def\etal{{\rm et al. }}
\def\kms{{\rm km \, s^{-1}}}
\def\Mpc{{\it h}^{-1}\, {\rm Mpc}}
\def\kpc{{\it h}^{-1}\, {\rm kpc}}
\def\Msol{{\it h}^{-1}\, {\rm M_\odot}}
\def\ie {{\rm i.e. }}
\def\eg {{\rm e.g. }}

\def\d {{\rm d}}

\def\lsim{\mathrel{\hbox{\rlap{\hbox{\lower4pt\hbox{$\sim$}}}\hbox{$<$}}}}
\def\gsim{\mathrel{\hbox{\rlap{\hbox{\lower4pt\hbox{$\sim$}}}\hbox{$>$}}}}

\begin{document}

\def\head{
\title{The Power Spectrum Dependence of Dark Matter Halo Concentrations}

\lefthead{Eke, Navarro \& Steinmetz}
\righthead{Power Spectrum and Dark Halo Concentration}

\author{Vincent R. Eke\altaffilmark{1}, Julio
F. Navarro\altaffilmark{2,3} and Matthias Steinmetz\altaffilmark{1,4}}

\affil{$^1$Steward Observatory, 933 N. Cherry Ave, Tucson, AZ 85721, USA}

\affil{$^2$Department of Physics and Astronomy, University of Victoria, 
Victoria, BC, V8P 1A1, Canada}

\affil{$^3$CIAR Scholar and Alfred P. Sloan Fellow}
\affil{$^4$David and Lucile Packard Fellow and Alfred P. Sloan
Fellow}
%\maketitle

\begin{abstract}

High-resolution N-body simulations are used to examine the power
spectrum dependence of the concentration of galaxy-sized dark matter
halos. It is found that dark halo concentrations depend on the amplitude
of mass fluctuations as well as on the ratio of power between small and virial
mass scales. This finding is consistent with the original results of
Navarro, Frenk \& White (NFW), and allows their model to be extended
to include power spectra substantially different from Cold Dark Matter
(CDM). In particular, the single-parameter model presented here fits the
concentration dependence on halo mass for truncated power spectra,
such as those expected in the warm dark matter (WDM) scenario, and
predicts a stronger redshift dependence for the concentration of CDM
halos than proposed by NFW. The latter conclusion confirms recent
suggestions by Bullock et al., although this new modeling differs from
theirs in detail. These findings imply that observational limits on the
concentration, such as those provided by estimates of the dark matter
content within individual galaxies, may be used to 
constrain the amplitude of mass fluctuations on galactic and
subgalactic scales. The constraints on $\Lambda$CDM models
posed by the dark mass within the solar circle in the Milky Way
and by the zero-point of the Tully-Fisher relation are revisited,
with the result that
neither dataset is clearly incompatible with the `concordance'
($\Omega_0=0.3$, $\Lambda_0=0.7$, $\sigma_8=0.9$) $\Lambda$CDM
cosmogony.  This conclusion differs from that reached recently by
Navarro \& Steinmetz, a disagreement that can be traced to
inconsistencies in the normalization of the $\Lambda$CDM power spectrum used in
that work.

\end{abstract}
\keywords{
cosmology ---
dark matter ---
galaxies: formation ---
galaxies: structure ---
}
}%%%end head
\twocolumn[\head]

\section{Introduction}\label{sec:intro}

Due to their large density, the central regions of dark matter
halos, where galaxies form according to the current paradigm of
structure formation, hold important astrophysical clues to the nature
of dark matter. This is why many studies have attempted to constrain
dark matter models on the basis of clues to the dark mass distribution
gained from detailed studies of the dynamics of gas and stars in
individual galaxies. The most straightforward method compares dark
mass distributions derived from rotation curves of disk galaxies with
detailed predictions of N-body simulations (Frenk \etal 1988; Flores
\etal 1993; Flores \& Primack 1994; Moore 1994; Moore \etal 199b), 
although similar
insight can be gained by inspecting the high-order moments of the
stellar velocity distribution in spheroid-dominated systems (Carollo
\etal 1995; Rix \etal 1997; Gerhard \etal 1998; Cretton \etal 2000;
Kronawitter \etal 2000).

Despite the simplicity of the rotation-curve method and the numerous
studies reported in the literature to date (see, e.g., Swaters 1999
for a comprehensive list of references), there is still no broad
consensus regarding the detailed distribution of dark matter in disk
galaxies, a situation that reflects the difficulties associated with
obtaining accurate circular velocities over a large dynamic range in
radius, as well as with accounting for the contribution of the
baryonic component to the rotation curve and for the uncertain
response of the dark material to the assembly of the galaxy. For
example, while constant-density `cores' in the dark mass
distribution appeared at first to be necessary to explain the rotation
curves of low surface brightness (LSB) dwarf galaxies (Flores \&
Primack 1994, Moore 1994, McGaugh \& de Blok 1998), the persuasiveness
of the observational evidence for these cores has recently been called
into question by careful reanalysis of the observational datasets (van
den Bosch \etal 2000; Swaters, Madore \& Trewhella 2000; van den Bosch
\& Swaters 2000).

At the same time, there is also considerable uncertainty in
theoretical predictions of the dark mass distribution at radii as
small as those probed by the rotation curve data. Most workers agree
that Cold Dark Matter (CDM) halos have density profiles that diverge
near the middle (a result that would be at odds with the alleged
cores of LSB dwarfs), but there is still controversy as to the exact
asymptotic behavior of the density near $r=0$. The work of Navarro,
Frenk \& White (1996, 1997) suggested that the central density may
diverge as fast as $r^{-1}$, but subsequent work has argued both for
steeper (e.g.  $r^{-1.4}$ in Moore \etal 1998) and shallower profiles
(e.g., $r^{-0.7}$ in Kravtsov \etal 1998, although it should be noted
that the authors have apparently now retracted this result, see Klypin
\etal 2000). Each of these models predicts, of course, quite different
dark matter contributions to disk galaxy rotation curves, making it
difficult to provide a sound interpretation of the observational
data. 

In other words, even if observations could constrain beyond the
dark mass distribution near the middle of disk galaxies, then there
would still be no consensus on the exact significance of that finding
for dark matter models.
The reasons for the disagreements in the theoretical predictions are
still being investigated, but in all probability they reflect the
inherent difficulties associated with simulating accurately and
reliably the dynamical behavior within individual galaxies, where the
density contrast exceeds $10^6$. Particles inhabiting these
regions go about their orbits thousands of times during a Hubble time,
making numerical results highly vulnerable to insidious systematic
artifacts associated with the choice of integrator, time-stepping, and
gravitational softening. Unfortunately a full account of the dependence of the
innermost density profiles of CDM halos on such numerical parameters
is still lacking, but the indication is that it will
require extreme care and a concerted numerical effort on massively
parallel computers to be able to characterize unequivocally the
behavior of the dark matter density profile within the regions probed
by rotation curve data.

Given the intrinsic difficulty in providing robust theoretical
predictions for the shape of the inner density profiles and the
unsettled status of the interpretation of current rotation curve
datasets, it is important to identify alternative observational and
theoretical comparison criteria that are less sensitive to numerical
and observational shortcomings. Navarro \& Steinmetz (2000a,b,
hereafter NS00a,b) have recently argued that one possible choice is to
use the total dark matter content within the main body of individual
spiral galaxies. 

The typical radii involved are of order $\sim 10$ kpc for a bright
spiral, which corresponds to about $3$-$5\%$ of the virial radii.
These regions are much less affected by numerical resolution issues
than the $\sim$kpc regions probed by rotation curves. Also, by
focusing on the total dark mass within this radius rather than on its
detailed radial distribution, both observational and theoretical
estimates are presumably more reliable. For example, as discussed by
NS00a, there are strict upper limits on the dark mass enclosed within
the solar circle in the Milky Way from detailed models of Galactic
dynamics (Dehnen \& Binney 1998; Gerhard 2000).  Such a constraint can
be extended to other spiral galaxies by examining the
zero-point of the Tully-Fisher (TF) relation. Indeed, provided that
stellar mass-to-light ratios and exponential scalelengths can be
estimated reliably, the TF relation allows for direct estimates of the
dark mass within a couple of exponential scalelengths from the middle
of the galaxy.

NS00a applied these constraints to a number of halos simulated within
the $\Lambda$CDM scenario, and concluded that the dark mass in
$\Lambda$CDM halos is too centrally concentrated to be consistent with
observations. This result added to an uncomfortably long list of
concerns regarding the viability of CDM on the scale of individual
galaxies, including the survival of a large number of small mass halos
within the virialized body of a parent halo (the `substructure'
problem, see Klypin \etal 1999; Moore \etal 1999a), as well as the
evidence for constant density cores in dark halos alluded to
above. Taken together, the evidence appeared to warrant a radical
revision of one or more of the premises of the CDM paradigm, and there
has been no shortage of proposals: self-interacting dark matter
(Spergel \& Steinhardt 2000), warm dark matter (Hogan \& Dalcanton
2000), fluid dark matter (Peebles 2000), fuzzy dark matter (Hu,
Barkana \& Gruzinov 2000), etc, all aim to provide a model that
behaves like CDM on large scales but with reduced substructure and
`concentration' on the scale of individual galactic halos.

If the results of NS00a,b hold and $\Lambda$CDM halos are too
concentrated to be consistent with observations, then what changes are
needed in order to reconcile the predictions of this scenario with
observations? Are changes in the overall normalization of the power
spectrum necessary, or does the {\it shape} of the $\Lambda$CDM
spectrum require modification? Do small-scale cutoffs in the power
spectrum (as expected in warm dark matter models) help? Or, in a more
general sense, what is the relationship between halo concentration and
the power spectrum of initial density fluctuations?

These are the questions addressed here through an extensive suite of
N-body simulations. A description of the numerical simulations is
given in Section~\ref{sec:meth}, including details of the various
power spectra chosen for this study. Section~\ref{sec:res} contains
the main results regarding the concentration of dark matter halos and
their dependence on the power spectrum, and Section~\ref{sec:comp}
uses these results to revisit the viability of the $\Lambda$CDM model
regarding the Milky Way and Tully-Fisher
constraints. Section~\ref{sec:sum} summarizes the main conclusions.

\section{Numerical Methods}\label{sec:meth}

\subsection{Cosmology and Power Spectra}

All of the simulations described here adopt the same cosmological
background model: a flat, cosmological constant-dominated universe
with matter density parameter $\Omega_0=0.3$, $\Lambda_0=0.7$,
and Hubble parameter $h=0.65$\footnote{The present value of Hubble's
constant is parameterized by $H_{0}=100 \, h~ \kms{\rm
Mpc}^{-1}$.}. Two different power spectrum shapes have been
considered. The first is the standard CDM spectrum, in the form given
by Bardeen \etal (1986), which is fully characterized by $\sigma_8$,
the present linear theory amplitude of mass fluctuations in spheres of
radius $8 \, \Mpc$, and by the value of the `shape' parameter,
$\Gamma$ (Bardeen \etal 1986, Sugiyama 1995).

\begin{figure}
\epsfxsize=\linewidth
\epsfbox{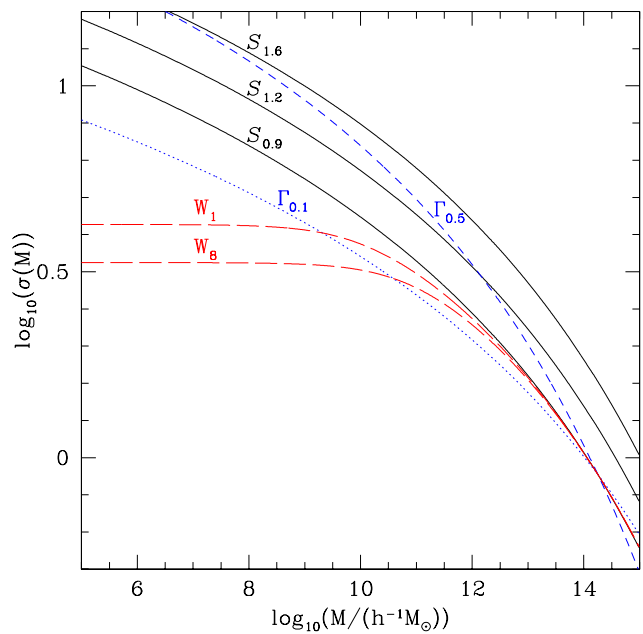}
\caption{
The $z=0$ linear amplitude of mass
fluctuations for the seven models investigated in this study. 
$\sigma(M)$ is calculated using a top-hat real-space window function. Solid
lines represent, from bottom to top, $\Lambda$CDM models $S_{0.9}$,
$S_{1.2}$ and $S_{1.6}$. The short-dashed line corresponds to
$\Gamma_{0.5}$ and the dotted line to $\Gamma_{0.1}$. The two WDM
models are shown with long-dashed lines; the top and bottom lines
correspond to models $W_1$ and $W_8$ respectively.}
\label{fig:sigmam}
\end{figure}

\begin{table}
\begin{center}
\caption{Power spectrum parameter choices.{\label{tab:pspec}}}
\begin{tabular}{llll}
\\ \hline 
Label & $\sigma_8$ & $\Gamma$ & $M_{f}/(10^{10} \Msol)$ \\
\hline 
$S_{0.9}$ & $0.9$ & $0.2$ & ~~~~~~~~~~~- \\
$S_{1.2}$ & $1.2$ & $0.2$ & ~~~~~~~~~~~- \\
$S_{1.6}$ & $1.6$ & $0.2$ & ~~~~~~~~~~~- \\
$\Gamma_{0.1}$ & $0.9$ & $0.1$ & ~~~~~~~~~~~- \\
$\Gamma_{0.5}$ & $0.9$ & $0.5$ & ~~~~~~~~~~~- \\
$W_1$ & $0.9$ & $0.2$ & ~~~~~~~~~~~1 \\
$W_8$ & $0.9$ & $0.2$ & ~~~~~~~~~~~8 \\
\hline
\end{tabular}
\end{center}
\end{table}

The second power spectrum shape aims to mimic a warm dark matter (WDM)
power spectrum: it is identical to the CDM spectrum on large scales
but its power is reduced on scales smaller than that of a
characteristic free-streaming mass, $P_{\rm WDM}(k)=P_{\rm CDM}(k)
\exp(-kR_f-(kR_f)^2))$, where $R_f$ is the comoving free-streaming
scale. Following Sommer-Larsen \& Dolgov (2000) and Avila-Reese \etal
(2000), a free-streaming wavenumber, $k_f$, is defined as that where
the WDM power spectrum is half the value for CDM. This implies $k_f
\approx 0.46/R_f$. The free-streaming mass is defined as
\begin{equation}
M_{f}=\frac{4\pi}{3} \bar{\rho}_{\rm WDM} \left(\frac{\lambda_f}{2}\right)^3,
\label{mfwdm1}
\end{equation}
with $\lambda_f=2\pi/k_f$ and $\bar{\rho}_{\rm WDM}$ being the density
of WDM. Expressing the free-streaming mass in terms of the
free-streaming scale yields
\begin{equation}
M_{f}=3.7\times10^{14} \, \Omega_{0} \, (R_f/ \Mpc)^3 \, \Msol.
\label{mfwdm2}
\end{equation}
This approximation to the actual WDM
cosmogony neglects the non-zero velocity dispersion of the warm dark
matter particle candidates, but recent work indicates that this
omission should have negligible consequences for the quantities of
interest here (Avila-Reese \etal 2000; Bode, Ostriker \& Turok 2000). 
On the other
hand, one advantage of this approximation is that the only difference
between the CDM and WDM runs is the small-scale behavior of the power
spectrum, which implies that systematic trends of halo structure with
power spectrum shape are easier to identify.

Table \ref{tab:pspec} contains a list of the specific parameters
chosen for the various models, and Figure \ref{fig:sigmam} shows
$\sigma(M)$, the $z=0$ amplitude of linear mass fluctuations in
spheres of a given mass corresponding to each power spectrum. In
total, seven different models were investigated; five $\Lambda$CDM
models with different parameter choices for $\sigma_8$ and $\Gamma$
and two WDM models with different free-streaming masses, $M_{f}$.
Model $S_{0.9}$ will be referred to hereafter as the `fiducial' model,
because it is roughly consistent with the local abundance of galaxy
clusters (Eke \etal 1996) and with the amplitude of CMB fluctuations
(Stompor, Gorski \& Banday 1995; Liddle \etal 1996). While a value of
$\Gamma=0.2$ was adopted for this default model, it is worth noting
that, according to the fit of Sugiyama (1995), $\Gamma=\Omega_0 \, h
\, \exp(-\Omega_b-(h/0.5)^{1/2} \, (\Omega_b/\Omega_0))$, 
and therefore $\Gamma \approx 0.16$ would be a
more appropriate value for the high baryon density parameter,
$\Omega_{\rm b} \approx 0.019 \, h^{-2}=0.045$ (for $h=0.65$),
advocated by Tytler \etal (2000).

\subsection{The Simulations}

For each model listed in Table \ref{tab:pspec}, the AP$^3$M code
(Couchman 1991) was used to evolve $128^3$ dark matter particles in a
$32.5 \, \Mpc$ cube from $z=24$ to $z=0$ using 2000 equal steps in
expansion factor. At $z=0$, four halos with circular velocities
between $180$ and $230 \, \kms$ (similar to that of the Milky Way)
were selected for resimulation from a list of halos identified by the
spherical overdensity group-finding algorithm (Lacey \& Cole
1994). Unless otherwise specified, halo circular velocities,
$V_{\Delta}$, are measured at the virial radius, $r_{\Delta}$; the
radius of a sphere containing a mean density $\Delta$ times the
critical value. The parameter $\Delta$ depends on $\Omega$ and
$\Lambda$ according to (\eg Eke, Navarro \& Frenk 1998),
\begin{equation}
\Delta(\Omega,\Lambda)= 178 \cases{ \Omega^{0.30},& if
$\Lambda=0$;\cr \Omega^{0.45},& if $\Omega + \Lambda =1$\cr}
\label{delta}
\end{equation}
and is $\approx 100$ at $z=0$ for the cosmology adopted here. In
addition to the circular velocity cuts, a criterion of relative
isolation was also enforced, so that halos considered for resimulation
were restricted to those without neighbors more massive than
$2.7\times 10^{11} \Msol$ within $1.5 \, \Mpc$. This selection
criterion increases the likelihood that the selected halos are close
to equilibrium, simplifying the interpretation of the results.
Besides the four `Milky Way' halos selected for each cosmogony,
further halos extending to circular velocities of order $100 \, \kms$
were also selected for resimulation in the fiducial $S_{0.9}$ model
and the WDM models.

The resimulations were performed using a multiple time-step N-body code
based on the algorithm described by Navarro \& White (1993), modified
to take advantage of the GRAPE3 hardware (Sugimoto \etal
1990). Particles were allowed to take up to $86,000$ time-steps during
their evolution from the starting redshift of $50$ to $z=0$. Each halo
has between $35,000$ and $85,000$ particles within the virial radius
at the final time. A Plummer gravitational softening of $\epsilon=0.6$
kpc was used in all resimulations of `Milky Way' halos. The extra
resimulated halos with $V_\Delta < 160 \,\kms$ were run using
$\epsilon=0.4$ kpc. A few simulations were rerun varying the numbers
of particles, and indicate that this numerical setup is appropriate for
making reliable measurements of the total mass within $5$-$10$ kpc.

\section{Power Spectrum and Halo Concentration}\label{sec:res}

\subsection{CDM and WDM density profiles}

\begin{figure}
\epsfxsize=\linewidth
\epsfbox{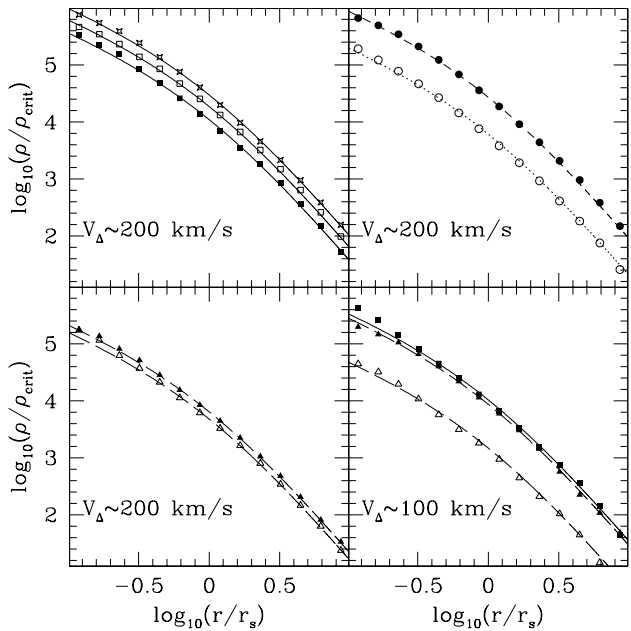}
\caption{ Density profiles of dark matter halos formed in the
different cosmological models.  Different symbols correspond to
different models, as follows: filled squares ($S_{0.9}$), stars
($S_{1.6}$), open squares ($S_{1.2}$), open circles ($\Gamma_{0.1}$),
filled circles ($\Gamma_{0.5}$), filled triangles ($W_1$), and open
triangles ($W_8$). Profiles shown are averages over the four `Milky
Way' halos ($V_{\Delta} \sim 200$ km/s) resimulated for each
cosmogony. A second average halo is also shown in the lower right
panel for models $S_{0.9}$, $W_1$ and $W_8$ corresponding to
$V_{\Delta} \sim 100$ km/s. The curves show NFW profiles fitted to the
average profiles. Line types are as in Figure \ref{fig:sigmam}.}
\label{fig:avdprof}
\end{figure}
 
Figure \ref{fig:avdprof} shows the density profiles at $z=0$
corresponding to the cosmologies listed in Table \ref{tab:pspec}.
Each profile is an average over the four `Milky Way' halos (i.e.,
halos with $V_{\Delta}$ in the range $180$-$230$ km/s). The mean
profile for each model is shown, together with fits of the form
proposed by Navarro, Frenk \& White (1996, 1997, hereafter NFW),
\begin{equation}
{\rho(r) \over \rho_{\rm crit}} = {\delta_c \over (r/r_s)(1+r/r_s)^2},
\label{nfw}
\end{equation}
where $\rho_{\rm crit}=3H^2/8\pi G$ is the critical density for
closure, $\delta_c$ is a characteristic density contrast, and $r_s$ is
a scale radius that corresponds to the region where the logarithmic
slope of the density equals the isothermal value,
d$\ln(\rho)/$d$\ln(r)=-2$. 

The main point to note here is that the NFW fitting formula works
quite well for $\Lambda$CDM halos in the radial range $0.1$-$10~
r_s$, in agreement with the results of NFW. This fitting
formula also reproduces the density profiles of WDM halos, even for
mass scales well below the free-streaming mass, $M_f$. This result has
been noted before (Huss, Jain \& Steinmetz 1999; Avila-Reese \etal 2000; Bode
\etal 2000), and allows the characterization of each halo
by two simple parameters: the mass inside $r_{\Delta}$ (the virial
mass $M_{\Delta}$) and the `concentration'
$c_{\Delta}=r_{\Delta}/r_s$. The concentration is directly related to
the NFW characteristic density contrast by
\begin{equation}
{\delta_c}={\Delta \over 3} { c_{\Delta}^3 \over
[\ln(1+c_{\Delta})-c_{\Delta}/(1+c_{\Delta})]},
\label{nfwdel}
\end{equation}
so that either parameter describes fully the density structure of a
halo of a given mass. In what follows, $c_{\Delta}$ will be adopted
except in the comparison with the results of NFW, where $\delta_c$
will be used. This is motivated by the fact that NFW adopted
$\Delta=200$ in their work, whereas the more general $\Delta$
definition of eq. \ref{delta} is adopted here. Note that $\delta_c$ is
independent of $\Delta$, but that concentration is not, so that one
should be careful when comparing concentration values quoted by
different authors. For the model considered here, $\Delta \approx 100$
at $z=0$, and $c_{\Delta} \sim 1.3~c_{200}$. Note that, although the
resolution of these simulations is good enough to measure
concentrations in a robust manner, it is not adequate to address the
ongoing controversy regarding the innermost slope of the density
profile.

\subsection{The Mass Dependence of Halo Concentration}\label{ssec:resm}

\begin{figure}
\epsfxsize=\linewidth
\epsfbox[130 20 450 750]{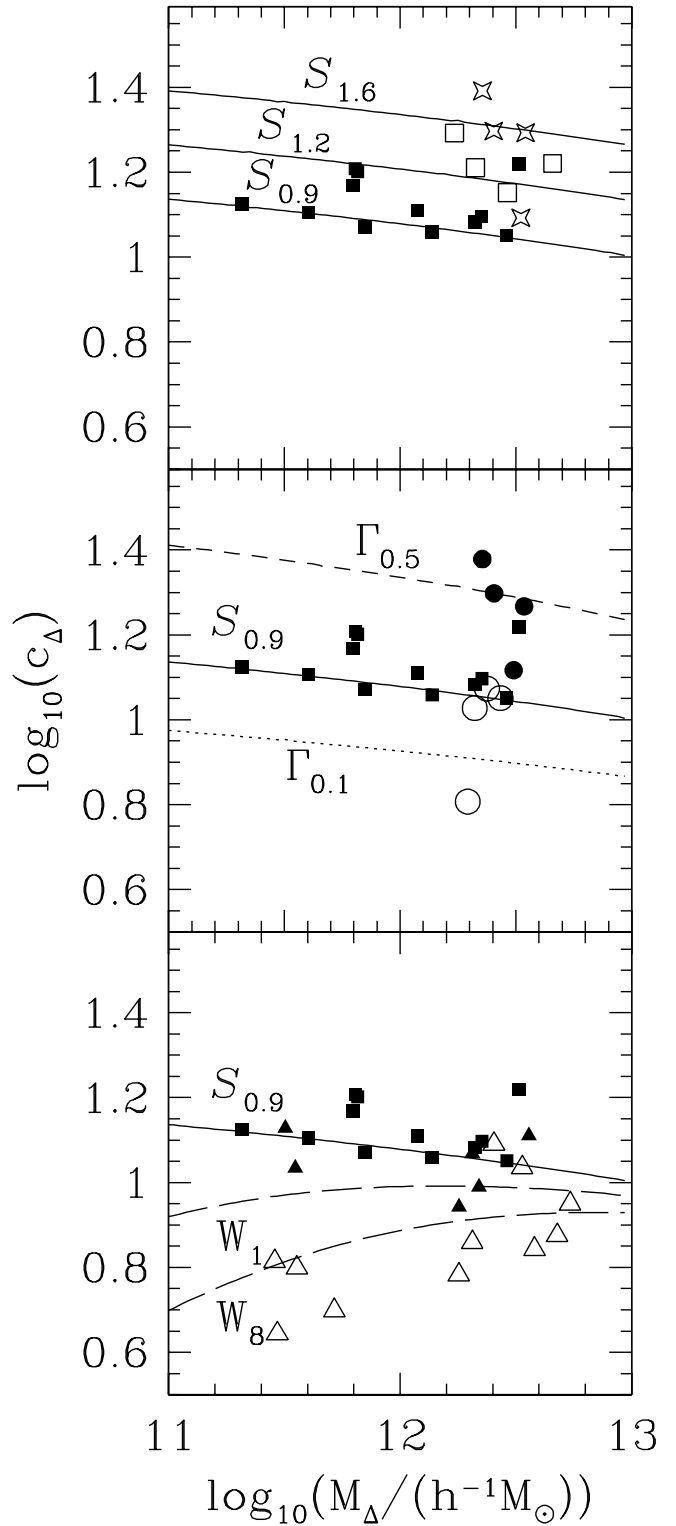}
\caption{ Concentration as a function of halo mass for all models,
split into three panels for ease of presentation. Lines in each panel
are fits to the data using the model described in \S \ref{ssec:model},
using the {\it same} value of $C_{\sigma}=28$ (the only free parameter
in the modeling) for {\it all} models. The top panel shows, from top
to bottom, $\Lambda$CDM models $S_{1.6}$ (starred symbols), $S_{1.2}$
(open squares), and the fiducial model $S_{0.9}$ (filled squares). The
fiducial model is repeated in all panels for comparison. The middle
panel shows models $\Gamma_{0.5}$ (filled circles), and $\Gamma_{0.1}$
(open circles). The bottom panel shows $W_1$ (filled triangles) and
$W_8$ (open triangles).  All line types are as in Figure
\ref{fig:sigmam}.}
\label{fig:sigint}
\end{figure}

Figure \ref{fig:sigint} shows the concentrations measured in the
simulations at $z=0$, as a function of the virial mass of each
halo. Different symbols correspond to different cosmogonies, as
described in the caption to Figure \ref{fig:avdprof}. The top panel
corresponds to the three $\Lambda$CDM models, $S_{1.6}$, $S_{1.2}$,
and $S_{0.9}$, from top to bottom, respectively. The middle panel
shows models $\Gamma_{0.5}$ and $\Gamma_{0.1}$, while the bottom panel
presents results corresponding to the warm dark matter models $W_1$
and $W_8$. Results for the fiducial model $S_{0.9}$ are repeated in
all panels.

There are a few things to note in this figure. Firstly, $\Lambda$CDM
concentrations increase with increasing $\sigma_8$ and decrease with
increasing mass. These trends are consistent with those reported by
NFW on the basis of lower resolution simulations, and support NFW's
interpretation that the concentration, or equivalently, the
characteristic density of a halo, reflects the mean density of the
universe at a suitably defined collapse time. Collapse redshifts
increase for higher values of the normalization parameter $\sigma_8$,
and are higher for low mass systems, reflecting the hierarchical
development of structure in CDM universes.  

\begin{figure*}
\epsfxsize=\linewidth
\epsfbox{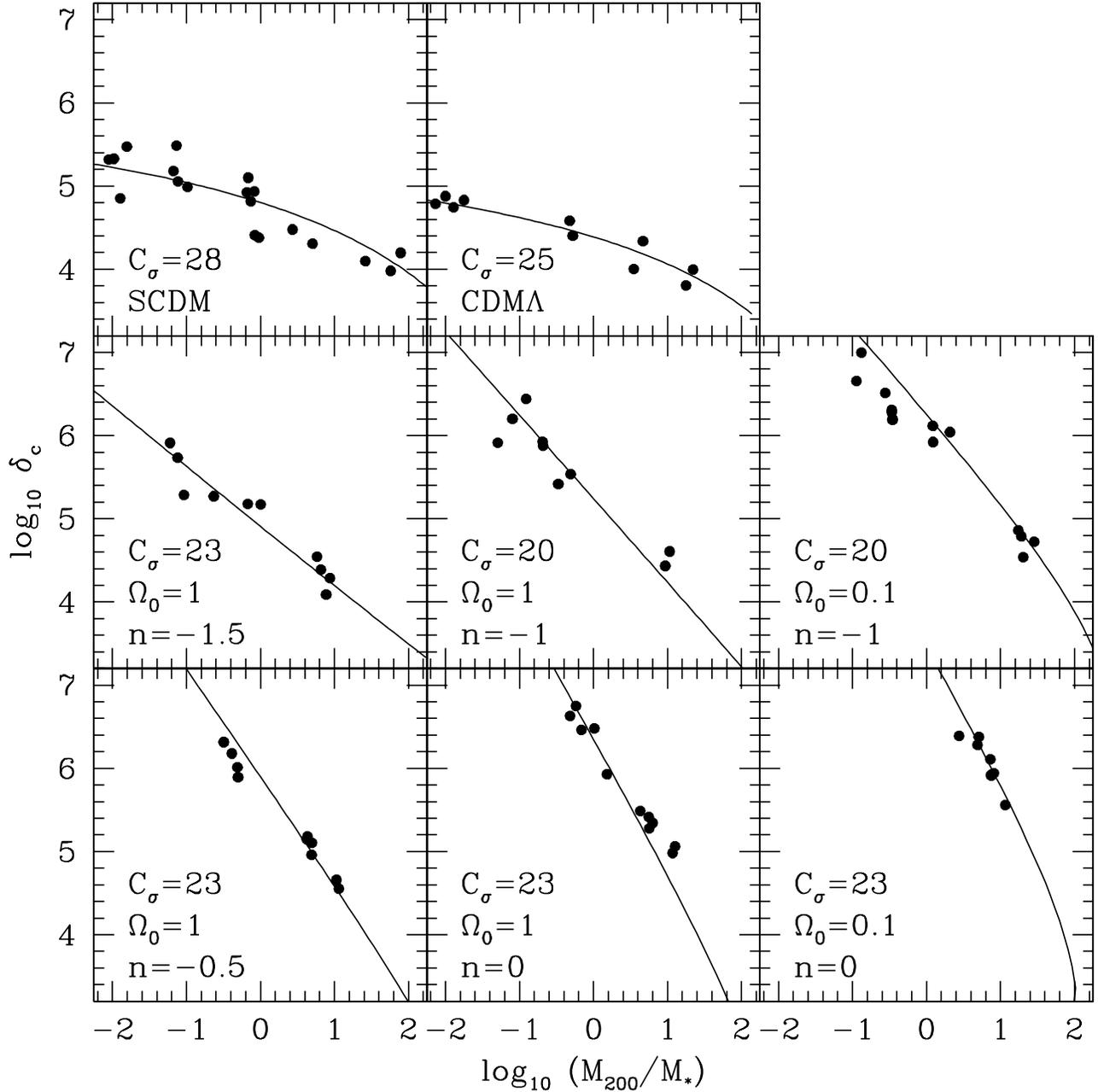}
\caption{ Mass dependence of halo characteristic densities, as
reported by Navarro, Frenk \& White (1997) (solid circles), compared
with the results of the model described in \S \ref{ssec:model}.
Constant $C_{\sigma}$ in eq. \ref{us1} has been chosen in each case so
as to provide a good fit to the simulation results at $M_{200} \approx
M_*$. Model SCDM
corresponds to the former `standard' CDM ($\Omega_0=1$, $\sigma_8=0.63$,
$\Gamma=0.5$). Model CDM$\Lambda$ has $\Omega_0=0.25$,
$\Lambda_0=0.75$, $\Gamma=0.19$, and $\sigma_8=1.3$. The rest
of the panels correspond to power-law power spectra, $P(k) \propto
k^n$; the value of $n$ and $\Omega_0$ is listed in each panel. Masses
are normalized to the characteristic clustering mass $M_*$, defined so
that $\sigma(M_*)=\delta_{\rm crit}$ (=1.686 for $\Omega_0=1$). This
corresponds to $M_*=1.6 \times 10^{13} h^{-1} M_{\odot}$ for SCDM and
$M_*=4.1 \times 10^{13} h^{-1} M_{\odot}$ for CDM$\Lambda$. Note that
excellent fits can be obtained in all cases with similar values of the
single free parameter $C_{\sigma}$.}
\label{fig:nfwfig5}
\end{figure*}

Secondly, $\Lambda$CDM concentrations depend very weakly on mass for
the range considered here; changing by only about $50\%$ over two decades in
mass for model $S_{0.9}$. 
Figure \ref{fig:nfwfig5} shows the simulation results presented by NFW
for a variety of different cosmological models and dark matter power
spectra, $P(k)$. The weak dependence on mass for the CDM models is surprising
when compared with the stronger trends observed for the
power-law power spectra simulations, labeled with the
spectral index $n$, where $P(k) \propto k^n$. As $n$ becomes more
negative, the concentration depends more weakly upon mass.
This is to be expected, since, as pointed out by NFW, the scaling
between $\delta_c$ and $M_{\Delta}$ found in their numerical simulation
is $\delta_c \propto M_{\Delta}^{-(n+3)/2}$, the same that links the
characteristic non-linear mass $M_{*}(z)$ and the mean cosmic density at
redshift $z$.\footnote{The characteristic clustering mass $M_{*}$ is
defined so that $\sigma(M_*)D(z)=\delta_{\rm crit}$ ($=1.686$ for
$\Omega=1$, consult Lacey and Cole (1993) and Eke, Cole \& Frenk (1996)
for other values of $\Omega$ and $\Lambda$).}
However, as can be readily seen in Figure \ref{fig:nfwfig5}, the
$\delta_c$-$M$ dependence found for CDM models is actually much weaker
than expected for $n \sim -1.5$, the `effective' CDM spectral index on the
mass scales probed by the NFW simulations. A more negative spectral
index seems necessary to explain the CDM results. This led NFW to
postulate that it is the amplitude of fluctuations on mass scales {\it
much smaller} than the virial mass that determine the
concentration. Consequently they introduced a (rather arbitrary)
parameter of order $\lsim 1\%$ (see parameter $f$ in equation
\ref{nfwps} below) in their modeling, in order to shift the mass scale
under consideration and reproduce the numerical results. This is a
rather unsatisfactory aspect of their modeling that lacks clear
interpretation. 

Finally, further clues can be gleaned from the concentration of
$\Gamma_{0.5}$ and $\Gamma_{0.1}$ halos.  $\Gamma_{0.1}$
concentrations are lower than $S_{0.9}$, which is not surprising given
that the amplitude of mass fluctuations is significantly lower on
galactic scales (Figure
\ref{fig:sigmam}). On the other hand, $\Gamma_{0.5}$ concentrations
are as high as $S_{1.6}$, although $\sigma(M)$ is in this case lower
than $S_{1.6}$ on galaxy-mass scales (Figure \ref{fig:sigmam}). This
again hints that the amplitude on virial mass scales is a poor
predictor of the concentration. These hints are confirmed by the
results of WDM model $W_8$, which shows a clear {\it reversal} of the
concentration versus mass trend on scales below a few times the free-streaming
mass $M_f$. $W_8$ concentrations {\it decrease} with decreasing mass
despite the fact that WDM $\sigma(M)$ increases towards low masses
before saturating at $M \ll M_f$ (Figure \ref{fig:sigmam}).

\subsection{A model for the power spectrum dependence of the concentration}\label{ssec:model}

Although the model proposed by NFW captures many of the qualitative
trends shown in Figure \ref{fig:sigint} it suffers from two main
shortcomings: (i) it introduces two arbitrary parameters
whose interpretation remains unclear, and (ii) it predicts a
redshift dependence for the concentration that is weaker than found in
recent numerical simulations (Bullock \etal 2000).  Bullock \etal
propose an alternative prescription, also with two free parameters, 
that results in improved agreement
between the predicted redshift dependence of concentrations and the
results of the numerical simulations. Their model follows NFW in
associating a halo's characteristic density with the average
background density at collapse time, but differs from NFW in the
definition of characteristic density and collapse time.

More specifically, NFW take the characteristic density of a halo to be
$\delta_c$ (see eq. \ref{nfwdel}), and use a constant of
proportionality, $C$, to relate this to the mean background density at
the collapse redshift, $z_c$, according to
\begin{equation}
\delta_c = C \, \Omega_0 \, (1+z_c)^3.
\label{nfw1}
\end{equation}
The collapse time is defined as that when,
according to the Press \& Schechter approach (Press \& Schechter 1974;
Lacey \& Cole 1993) , half the
virial mass of the halo was first contained in progenitors more
massive than a fraction $f$ of the final mass. This implies that
\begin{equation}
{\rm erfc} \left\{ \frac{\delta_{\rm crit}(z_c)-\delta_{\rm crit}(z_o)}
{\sqrt{2[\sigma^2(fM)-\sigma^2(M)]}} \right\} = \frac{1}{2},
\label{nfwps}
\end{equation}
where $z_o$ denotes the redshift at which the halo is identified,
$\delta_{\rm crit}(z)=\delta_{\rm crit}(0)/D(z)$ is the spherical
top-hat model linear overdensity threshold and $D(z)$
represents the linear theory growth
factor. ($D(z)=(1+z)^{-1}$ if $\Omega_0=1$, $\Lambda_0=0$ is
conventionally normalized to unity at $z=0$; formulae for other values
of $\Omega_0$ and $\Lambda_0$ can be found in Peebles 1980.)
For their simulations, NFW found a good fit by adjusting the two free
parameters to be $C=3000$ and $f=0.01$.

Bullock et al., on the other
hand, choose the characteristic density, ${\tilde \rho}_s$, to be such
that
\begin{equation}
M_{\Delta}={4\pi \over 3} r_s^3 \, {\tilde \rho}_s,
\label{bull1}
\end{equation}
and specify the collapse redshift, $z_c$, solely in terms of
$\sigma(M)$, so that
\begin{equation}
D(z_c) \, \sigma(F\, M_{\Delta})=1.686,
\label{bull2}
\end{equation}
where $F=0.01$. Their
second free parameter, $K$, relates the characteristic density to the
background density via
\begin{equation}
{\tilde \rho}_s = K^3 \Delta(z_o) \rho_{\rm crit}(z_c)
\label{bull2.5}
\end{equation}
and feeds through into the concentration as
\begin{equation}
c_\Delta=K \left(\frac{1+z_c}{1+z_o}\right).
\label{bull3}
\end{equation}
$K=4$ provides a good fit to their $\Lambda$CDM simulation results.

This model, like that of NFW, suffers from the introduction of two
arbitrary parameters ($F$ and $K$) whose interpretation remains unclear.
Furthermore, the definition of collapse epoch given in equation
\ref{bull2} implies that for a truncated power spectrum such as
WDM, halo concentrations will still increase monotonically with decreasing
mass, approaching a constant at $M \ll M_f$. 
This is at odds with the results presented in the previous
section (see also Bode \etal 2000), which show that WDM halo
concentrations {\it decrease} on mass scales below a few times the
free-streaming mass. These results strongly suggest that it is not
only the amplitude of the power spectrum, but also its shape, that
determine the concentration of dark matter halos. In particular, only
a modeling that includes such shape dependence will be able to
reproduce the somewhat counterintuitive dependence of concentration on
halo mass found for truncated power spectra such as $W_8$ (see Figure
\ref{fig:sigint}). 

After some experimentation, a simple model has been produced that
matches the mass dependence of halo concentrations for the simulations
presented here. Furthermore, the
same model also fits all of the original NFW results, whilst
modifying the redshift dependence of concentrations so that they are
compatible with the recent results of Bullock et al. The new model has
{\it a single free parameter} and is of more general applicability,
since it can be applied to truncated power spectra, where Bullock et
al.'s prescription fails. This model also removes the need for the
arbitrary, small mass fraction constant introduced by NFW and Bullock
\etal (f=0.01 in equation \ref{nfwps} and F=0.01 in equation \ref{bull2})
by postulating that the concentration of
a halo is controlled by a combination of the amplitude {\it and} shape
of the power spectrum. 

Consider the `effective' amplitude of the power
spectrum on scale $M$, defined by,
\begin{equation}
\sigma_{\rm eff}(M)=\sigma(M) \, \left(-{\d \ln(\sigma) \over
\d \ln(M)}(M)\right).
\label{sigeff}
\end{equation}
This effective amplitude modulates $\sigma(M)$ so that, for WDM-like
spectra, it {\it decreases} on mass scales smaller than a few times the
free-streaming mass $M_f$. 
In broad terms, a given mass scale $M$ collapses when $D(z) \sigma(M)$
is at least unity. This time is controlled by the redshift evolution
of the linear growth factor, $D(z)$, appropriate for the cosmological
model under consideration.  Following this, the collapse redshift,
$z_c$, of a halo of mass $M$ may be identified as
\begin{equation}
D(z_c) \sigma_{\rm eff}(M_s)= {1 \over C_{\sigma}}
\label{us1}
\end{equation}
where $C_{\sigma}$ is a constant and $M_s$ is the mass contained
within $r_{\rm max}=2.17 \, r_s$, the radius at which the circular
velocity of an NFW halo reaches its maximum. The requirement for
collapse that
$D(z_c) \sigma(M_s) \geq 1$ implies that $C_\sigma \geq
-\d\ln\sigma/\d\ln M$. For a power-law fluctuation spectrum with $P(k)
\propto k^n$ then this yields $C_\sigma \geq 6/(n+3)$.
As in the models of NFW
and Bullock et al., the mean density of the universe at the collapse
redshift can then be used to calculate a characteristic density for
the halo. Defining the characteristic density of the halo to be, as in
Bullock \etal (see eq. \ref{bull1}),
\begin{equation}
{\tilde \rho}_s=\Delta(z_o) \, \rho_{\rm crit}(z_o) \, c_\Delta^3, 
\label{ourrhos}
\end{equation}
and setting this to equal the spherical collapse top-hat density
at the collapse epoch, $\rho_{\rm sc}$, where
\begin{equation}
\rho_{\rm sc}(z_c)=\Delta(z_c)\rho_{\rm crit}(z_c)=\Delta(z_c)
\frac{\bar{\rho}_0 (1+z_c)^3}{\Omega(z_c)}
\label{rhosc}
\end{equation}
yields,
\begin{equation}
c_\Delta^3= {\Delta(z_c) \over \Delta(z_o)} \, {\Omega(z_o) \over \Omega(z_c)} \,
\left({1+z_c \over 1+z_o}\right)^3.
\label{us2}
\end{equation}
Equations \ref{us1} and \ref{us2} describe the concentration of a halo of
given mass, once the single free parameter in eq. \ref{us1},
$C_{\sigma}$, has been specified. As the characteristic mass scale at
which the effective amplitude of the power spectrum is evaluated
depends on $r_s$, and therefore on $c_\Delta$, 
equations \ref{us1} and \ref{us2} need to be solved iteratively to 
yield the combination of $c_\Delta$ and $z_c$ \footnote{An algorithm to
perform this calculation for CDM and WDM power spectra is available on
request from the authors.}.

This model reproduces, with roughly the same value of $C_{\sigma}$,
the results of the simulations presented here, all of the original
results of NFW, as well as the redshift dependence advocated by
Bullock et al. This is shown by the curves in Figure \ref{fig:sigint},
which show the result of applying the model, at $z=0$, to the seven
cosmogonies adopted in this study. Solid line types are used for the
$S$ models, short-dashed and dotted lines for $\Gamma_{0.5}$ and
$\Gamma_{0.1}$, respectively, while long-dashed lines are used for
WDM. All of the curves use the same value for the proportionality
constant in eq. \ref{us1}, $C_{\sigma}=28$. The model reproduces very
well the trends with mass, normalization, and shape of the power
spectrum seen here, including the counterintuitive trend towards lower
concentrations seen in the low-mass $W_8$ halos. Table \ref{tab:eg} contains
a list of concentrations for a $10^{12} \Msol$ halo identified at $z=0$
in a variety of commonly studied cosmological models. This illustrates
the interplay between $\Omega_0$, $\Lambda_0$, $\sigma_8$ and $\Gamma$.

\begin{table}
\begin{center}
\caption{Values of $c_\Delta$ and $c_{200}$ 
for a $10^{12} \Msol$ halo identified at
$z=0$ in four commonly studied cosmological models.{\label{tab:eg}}}
\begin{tabular}{lllllll}
\\ \hline 
$\Omega_0$ & $\Lambda_0$ & h & $\sigma_8$ & $\Gamma$ & $c_\Delta$ &
$c_{200}$ \\
\hline 
$1$ & $0$ & $0.5$ & $0.5$ & $0.5$ & $12.3$ & $11.7$ \\
$1$ & $0$ & $0.5$ & $0.5$ & $0.2$ & $6.7$ & $6.4$ \\
$0.3$ & $0$ & $0.65$ & $0.9$ & $0.2$ & $18.6$ & $15.2$ \\
$0.3$ & $0.7$ & $0.65$ & $0.9$ & $0.2$ & $12.0$ & $8.9$ \\
\hline
\end{tabular}
\end{center}
\end{table}

The model described above also reproduces the original results of NFW
quite well. This is shown in Figure \ref{fig:nfwfig5}, where the
density contrast $\delta_c$ is plotted as a function of the mass
enclosed within a $\Delta=200$ sphere, the parameters used by NFW. It
is apparent from this figure that the model also reproduces the results
of the eight cosmogonies studied by NFW, including open models with
$\Omega_0$ as low as $0.1$, again with approximately a single value of
the constant $C_{\sigma}$.

\subsection{The Redshift Dependence of Halo Concentration}

According to equations \ref{us1} and \ref{us2}, the model predicts
that, at fixed halo mass, in an Einstein-de Sitter cosmogony $c_\Delta(M,z)
\propto (1+z)$. This relation agrees with the prediction of the model
by Bullock \etal for the evolution of halo concentration.  However,
for low density universes the scaling with redshift is not the same as
theirs, and it is therefore important to verify that it is still in
good agreement with the numerical results.

\begin{figure*}
\epsfxsize=\linewidth
\epsfbox[0 20 630 750]{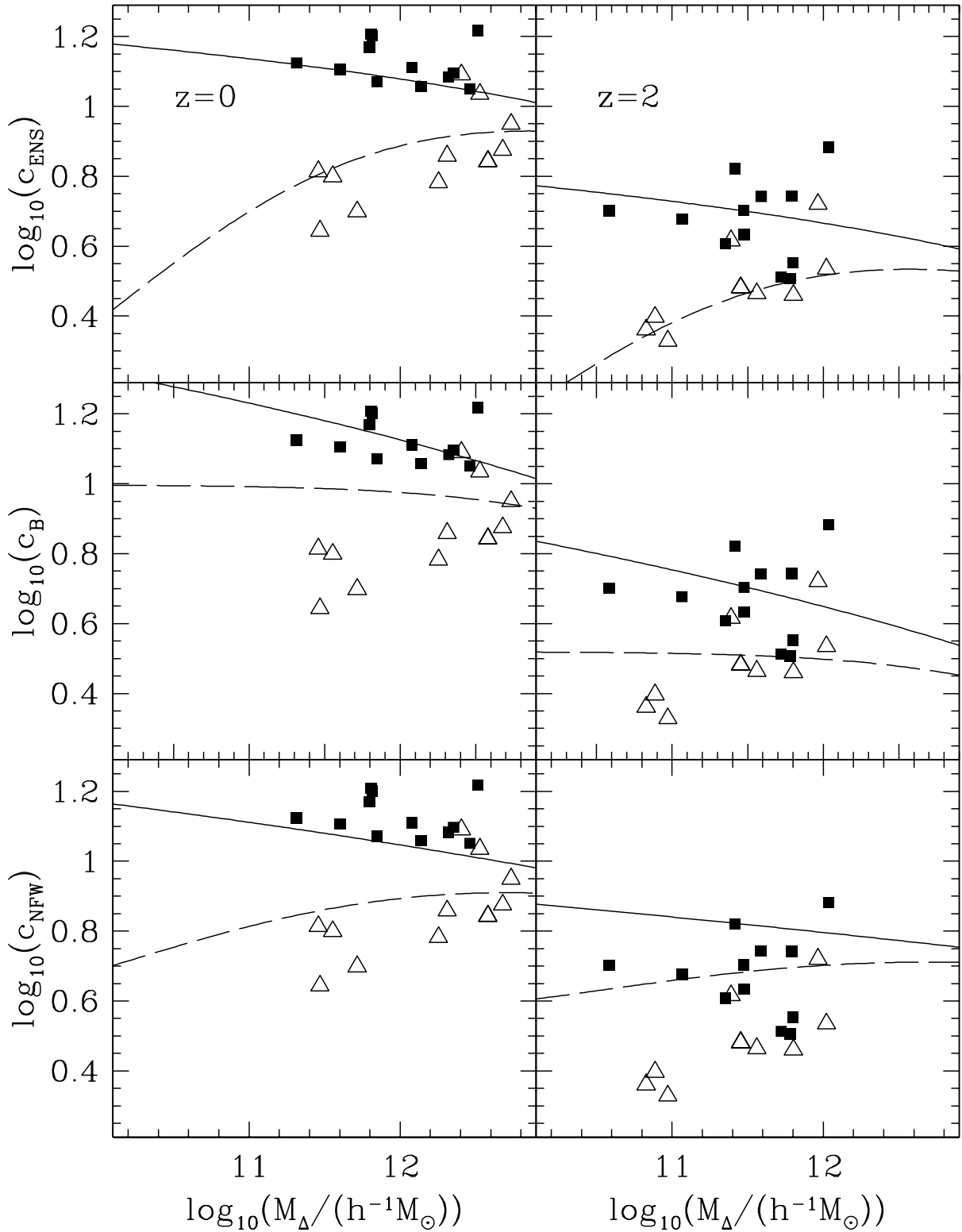}
\caption{ Comparison between concentrations measured at $z=0$ (left
panels) and $z=2$ (right panels), and the predictions of three
different models. `ENS' corresponds to the model presented in this
paper (top panels), `B' to that in Bullock et al (2000, middle
panels), and `NFW' to concentrations computed using the procedure
outlined in the Appendix of NFW (bottom panels). The filled-in squares
correspond to $S_{0.9}$ halos, whereas open triangles correspond to
$W_8$ halos. The solid and dashed curves show the model predictions for
$S_{0.9}$ and $W_8$, respectively.}
\label{fig:cmcomp}
\end{figure*}

Figure \ref{fig:cmcomp} compares the predictions of all three
different concentration models with the results of the numerical
simulations at $z=0$ and $2$. The comparison includes all halos in the
$S_{0.9}$ and $W_8$ simulations with more than $2500$ particles within
$r_\Delta$.
Concentrations labelled with an `ENS' subscript in the top row
correspond to the model presented here, `B' to Bullock et al.'s (middle
row), and `NFW' for the NFW model predictions in the bottom
panels. ENS concentrations use $C_{\sigma}=28$ in eq. \ref{us1}. `B'
concentrations use $F=0.01$ and $K=4$. NFW concentrations use $C=3000$
and $f=0.01$. The typical halo mass
range probed varies from $10^{11}$-$10^{13} \Msol$ at $z=0$ to
$10^{10.5}$-$10^{12} \Msol$ at $z=2$.

The top two panels show that the model presented here predicts a
redshift dependence in good agreement with the simulation results,
both for $S_{0.9}$ and $W_8$. The middle panels show that the Bullock
\etal model also fits the results of the fiducial $\Lambda$CDM
runs at $z=0$ and $z=2$, but that their model fails to capture
the mass dependence seen in the $W_8$ simulations.  This illustrates
the point that was made earlier that it is the effective
normalization (equation \ref{sigeff}), 
rather than simply $\sigma(M)$, that determines halo
concentrations. The bottom row highlights the weak redshift dependence
predicted by the NFW model compared with the simulation results, as
noted by Bullock et al.; it slightly under-predicts the fiducial model
concentrations at $z=0$ but over-predicts them at $z=2$.

In summary, the data in Figure \ref{fig:cmcomp} shows that the
redshift evolution predicted by the model presented here is consistent
with the simulation results. Even at $z=5$, the highest redshift with
simulation data in figure 11 of Bullock et al., the concentrations of
$10^{12} \Msol$ halos are $c_\Delta=2.6$ and $2.5$ for the model in Section
\ref{ssec:model} and that of Bullock \etal respectively. Thus,
it is not possible to discriminate between the slightly different redshift
dependences of these two models.

\section{Comparison with Observations}\label{sec:comp}
\subsection{$\Lambda$CDM and the Dark Mass within the Solar Circle}\label{ssec:mway}

Observations of star and gas kinematics provide well defined
constraints on the dark matter content of the Milky Way within the
solar circle, $R_0=8.5$ kpc. As discussed by NS00a (see that paper for
full references), a simple upper limit on the dark mass within $R_0$,
$M_{\rm dark}(r<R_0)$, may be obtained by combining the observed
circular velocity at the Sun's location, $V_c(R_0)=220 \, \kms$, with
estimates for the total mass and exponential scalelength of the
Galactic disk ($M_{\rm disk}=6 \times 10^{10} M_{\odot}$ and $r_{\rm
disk}=3.5$ kpc, respectively). The result (note that there is a typo
in eq.1 of NS00a),
\begin{equation}
M_{\rm dark}(r<R_0) \lsim 4.3 \times 10^{10} M_{\odot},
\label{mdarkway}
\end{equation}
constitutes an upper limit because the simple calculation described
above neglects two potentially important effects: (i) the contribution
of the bulge, and (ii) any potential contraction that the dark halo
may have experienced as a result of the assembly of the galaxy.

\begin{figure}
\epsfxsize=\linewidth
\epsfbox[130 20 450 750]{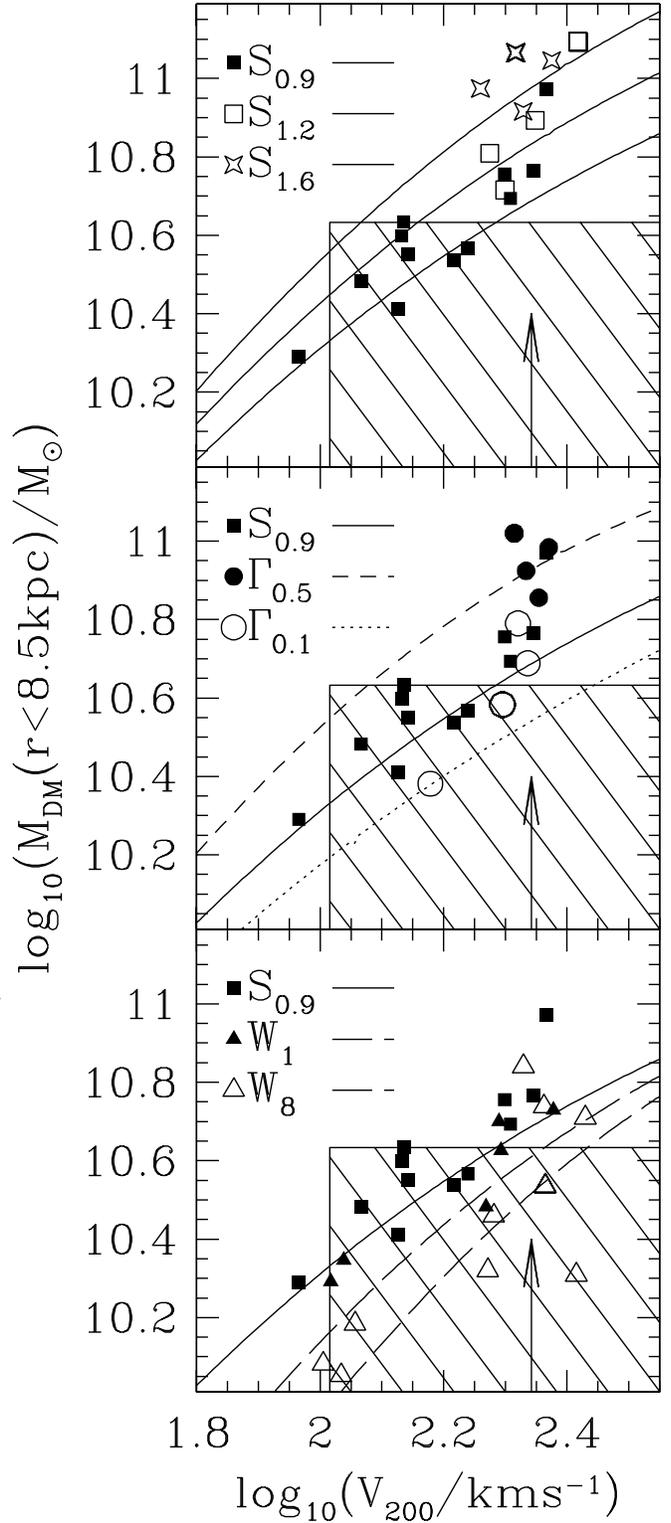}
\caption{ Dark matter halo masses within the solar circle ($8.5$ kpc)
compared with constraints derived from dynamical observations of the
Milky Way (hatched region). A vertical arrow marks a circular velocity
of $220~\kms$. The different symbols correspond to different
cosmogonies, as specified in Figure \ref{fig:avdprof}. Lines
correspond to NFW profiles assuming concentrations given by the model
in \S \ref{ssec:model}.}
\label{fig:mway}
\end{figure}

The constraint expressed in eq. \ref{mdarkway} is straightforward to
compare with the results of the numerical simulations described
here. This is done in Figure \ref{fig:mway}, where the dark mass
within $8.5$ kpc is plotted as a function of $V_{200}$, the halo
circular velocity for $\Delta=200$ used in NS00a, for all of the
halos. The top panel shows halos formed in the $S_{0.9}$, $S_{1.2}$,
and $S_{1.6}$ models, the three different normalizations chosen for
the $\Lambda$CDM scenario. At $V_{200}=220 \, \kms$ (marked with an
arrow in Figure \ref{fig:mway}), $M_{\rm dark}(r<R_0)$ increases
approximately in proportion to $\sigma_8$. In light of the modeling
described above, this can be attributed to the higher average collapse
times that result from the choice of higher normalizations.

The lines in this panel correspond to the mass within $8.5$~kpc
predicted by the model described in \S\ref{ssec:model}.  Combining
this model with the constraint in eq. \ref{mdarkway} it is possible to
estimate
the range of circular velocities allowed for the halo of the Milky
Way as a function of the normalization parameter $\sigma_8$:
\begin{equation}
\sigma_8 \, v_{220} \, m_{\rm dark} < 0.8,
\label{mwaycon}
\end{equation}
where $v_{220}$ is the circular velocity,
$V_{200}$, of the Milky Way halo in units of $220 \, \kms$, and
$m_{\rm dark}$ is $M_{\rm dark}/(4.3 \times 10^{10} M_\odot)$.  This
suggests that, for $\Gamma=0.2$, the circular velocity of the
halo of the Milky Way should be somewhat less than $220 \, \kms$,
unless $\sigma_8 < 0.9$ or the Milky Way halo has an unusually low
concentration for its mass. The fiducial $S_{0.9}$ model may be
reconciled with the Milky Way constraint if $V_{200} \lsim 195 \,
\kms$. A knock-on effect of moving the Milky Way into a smaller halo
would be that any predictions of the galaxy luminosity function made
by mapping mass into luminosity using the properties of the Milky Way
would find that Milky Way-type galaxies were more abundant than before.
As $M \propto V^3$, and assuming that the luminosity of a galaxy is
proportional to its mass, it only takes a $20\%$ decrease in halo
circular velocity to halve the luminosity. Thus, it
remains to be seen whether assigning the luminosity of the Milky Way
to the many halos with $V_{200} \approx 195 \,\kms$ is consistent with
the luminosity function of
bright spirals (see, e.g., Cole \etal 1994; Cole \etal 2000).

A lower bound on the mass of the Milky Way halo may be derived by
requiring that the total baryonic mass of the Galaxy does not exceed
the baryon mass within the virial radius of the halo. Assuming
$\Omega_b=0.019 \, h^{-2} \approx 0.045$, the minimum halo mass
corresponds to a circular velocity of $\sim 105 \, \kms$. This is
lower than the $130 \, \kms$ derived by NS00a, because of the lower baryon
fraction ($\Omega_b=0.0125 h^{-2}$) and slightly higher Hubble constant
($h=0.7$) adopted by those authors. 

From Figure \ref{fig:mway}, halos in the fiducial $S_{0.9}$
$\Lambda$CDM model with circular velocities in the range $(105,190) \,
\kms$ appear consistent with the Milky Way constraint. For
$\sigma_8=1.2$ the range of acceptable halo masses is narrower, and
essentially no halo agrees with the observational constraints if
$\sigma_8=1.6$.  This is reminiscent, although less stringent, than
the conclusion reached by NS00a, who argued that $\sigma_8=1.14$
$\Lambda$CDM halos were too concentrated to be consistent with this
constraint. However, a reanalysis of the NS00a dataset reveals that
because of inconsistencies in the normalization procedure for their
$\Lambda$CDM simulations
\footnote{This error originated in the fact that NS00a used the
transfer function proposed by Davis \etal (1985) to displace
particles, while normalizing the power spectrum using the value at the
Nyquist frequency of the original low-resolution simulation given by
the CDM transfer function fit of Bardeen \etal (1986). At this small
scale, the two fits give power spectrum values that differ by almost a
factor of two, and this led to a systematic discrepancy between actual
and intended normalizations. This error only affected the $\Lambda$CDM
models of NS00a,b. All other models, including NFW's, are free from
this problem.} 
, those authors had effectively
normalized their power spectra to $\sigma_8 \sim 1.6$ rather than
$\sigma_8=1.14$. After correcting for this error, the results in
NS00a,b are consistent with those reported here.

\subsection{The zero-point of the Tully-Fisher relation}\label{sec:tf}

As discussed by NS00a,b, the analysis of Section \ref{ssec:mway} 
can be extended to
other spiral galaxies by examining the correlation between galaxy
luminosity and the rotation speed of their gas and stars: the
Tully-Fisher relation (Tully \& Fisher 1977). Provided that stellar
mass-to-light ratios and exponential scalelengths can be estimated
reliably, it is possible to evaluate the disk contribution to the
circular velocity at $2.2$ exponential scalelengths (where the disk
contribution peaks and optical Tully-Fisher velocities are
typically measured) and derive constraints on the total dark mass
contained within this radius.

\begin{figure}
\epsfxsize=\linewidth
\epsfbox{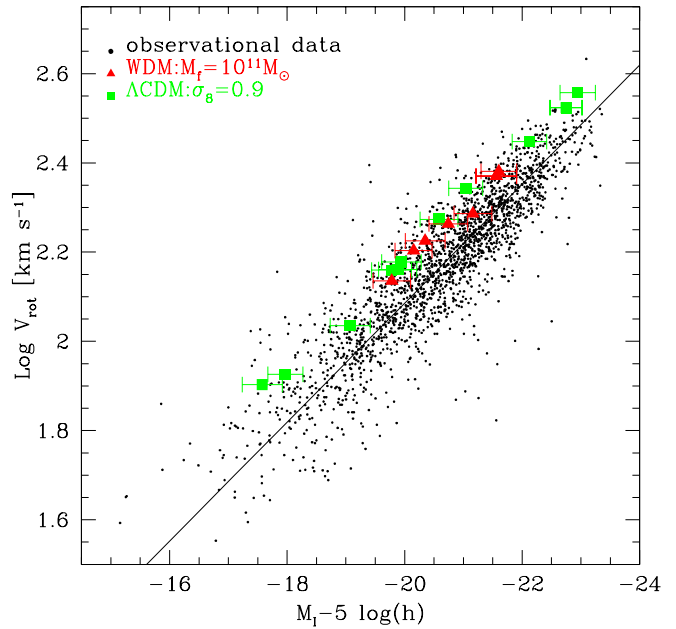}
\caption{ The I-band Tully-Fisher relation compared to the result of
numerical simulations in the fiducial $\sigma_8=0.9$ $\Lambda$CDM
(filled squares) and in the $W_1$ warm dark matter (filled triangles)
scenarios.  Dots are a compilation of the data by Giovanelli et al
(1997), Mathewson et al (1992), and Han \& Mould (1992). The solid
line is the best fit to the data advocated by Giovanelli et
al. Horizontal `error bars' in the simulation results span the range
in luminosities derived from assuming a Scalo or a Salpeter
IMF. Simulation circular velocities are measured at $r_{\rm gal}=20
(V_{200}/220 \, \kms) \, \kpc$. Note that the slope, scatter and
zero-point of the numerical TF relation are all in reasonable agreement
with observation.}
\label{fig:tf}
\end{figure}

The more concentrated a halo is, the faster a disk of given mass and
radial scale must rotate to attain centrifugal equilibrium. Thus, as
shown by NS00a,b, the zero-point of the Tully-Fisher relation provides
a direct constraint on halo concentrations. Although these authors
conclude that $\sigma_8=1.14$ $\Lambda$CDM halos are too concentrated
to be consistent with the I-band Tully-Fisher relation, as discussed
in \S \ref{ssec:mway} their conclusions were affected by an inconsistent
normalization of the power spectrum. Given that the simulations of
NS00a,b effectively probed a $\sigma_8\approx 1.6$ $\Lambda$CDM model
and concentration depends strongly on $\sigma_8$, it is appropriate to
revisit the issue and verify whether the fiducial $S_{0.9}$ model
is consistent with observations.

\subsubsection{Gasdynamical simulations}

To this aim, a number of gasdynamical simulations
including star formation and feedback have been run using GRAPESPH, a code that
combines the hardware N-body integrator GRAPE with the
smoothed-particle hydrodynamics (SPH) technique (Steinmetz 1996). The
simulation setup and analysis are identical to those described in
Navarro \& Steinmetz (1997) and Steinmetz \& Navarro (1999), and the
reader is referred
there for details.  In brief, the same initial conditions described
above for the dark matter-only runs are used with the addition of gas,
assuming a value of $\Omega_b=0.0125 h^{-2}$ for the baryon density
parameter. Models with gas have typically $20,000$ gas particles and
the same number of dark matter particles. Up to $5,000$ of these end
up in a galaxy at $z=0$, resulting in lower resolution than the
N-body simulations discussed in previous sections.  Gas particle
masses range from $4.5 \times 10^6 M_{\odot}$ to $2.5 \times 10^{8}
M_{\odot}$, depending on the model considered. 

Model galaxies are unmistakably identified in the runs as star and gas
clumps with high density contrast.  Only halos with more than $500$
dark particles within the virial radius have been retained for
analysis. The properties of the luminous component are computed within
a radius, $r_{\rm gal}=20 \, (V_{200}/220 \, \kms) \kpc$. This radius
contains all of the baryonic material associated with the galaxy and
is well outside the region compromised by numerical resolution
effects. All of the rotation speeds are also computed at that
radius. Note that $r_{\rm gal}$ exceeds the radii at which
Tully-Fisher velocities are typically measured, but given the lower
resolution of these simulations, rotation speeds at smaller radii are
quite uncertain. This comparison therefore assumes that the circular
velocity curves of actual disk galaxies remain approximately flat out
to $r_{\rm gal}$.

\subsubsection{The I-band TF relation}

Figure \ref{fig:tf} compares the observed I-band Tully-Fisher relation
(dots) with the numerical results for galaxies selected in the
fiducial $S_{0.9}$ $\Lambda$CDM model (filled squares) and in the
$W_1$ WDM model. There is reasonable agreement between observation and
simulations. The slope of the numerical TF relation is consistent with
the observed value, and the scatter is much smaller ($0.12$ mag rms
for $S_{0.9}$ and $0.10$ mag rms for $W_1$) than observed. These two
conclusions are in agreement with the results of NS00a,b.

The main difference with that work is that now even the zero-point of
the numerical relation appears to match reasonably well the observed
value; the zero-point offset between simulations and
observations is $\sim 0.5$ mag, compared with the $1.5$ mag offset
reported by NS00a,b. The reason for the discrepancy can again be
traced to the lower concentrations of $S_{0.9}$ halos compared with
the results of NS00a,b. \footnote{The
normalization problem only affected the $\Lambda$CDM runs in those
papers. All the results concerning the standard $\Omega=1$ CDM model
remain unchanged.} Figure \ref{fig:tf} also shows that there is little
difference in the TF results obtained for $W_1$ or $S_{0.9}$,
supporting the interpretation that the halo concentration is the main
factor responsible for the zero-point of the numerical TF relation.

The $0.5$ mag difference between simulation and observation is not too
worrying given that the simulated galaxies have colors that are
slightly too red compared with their TF counterparts. The average
$B-R$ color of the simulated galaxies is $1.2$, with little
dependence on luminosity. For comparison, the average $B-R$ in
Courteau's (1997) sample is $\sim 0.8$. This suggests that star
formation in the simulations occurs too early. Any modification to the
feedback algorithm that remedies this will also tend to make the
stellar population mix in the simulated galaxies brighter. If this
correction can bring the stellar I-band mass-to-light ratios down from
$2.5$ to $1.5$, a value more in keeping with the results of Bell \& de
Jong (2000), then the $0.5$ mag gap should be possible to bridge.

In summary, it appears that if the I-band stellar mass-to-light ratio
of TF galaxies is of order $(M/L)_I \approx 1.5$ then $\Lambda$CDM
halos are consistent with the slope, scatter and zero-point of the
I-band Tully-Fisher relation. Note however, that while halos formed in
the fiducial $\Lambda$CDM scenario appear to have concentrations
consistent with observational constraints, other problems associated with
the assembly of disk galaxies through merging persist. In particular,
the angular momentum (and size) of simulated disks is still quite
below observed values, again suggesting that perhaps the feedback
algorithm is not effective enough at preventing the early collapse of
baryons into protogalactic potential wells (Navarro \& Steinmetz
1997). Accounting simultaneously for the luminosity, velocity, and
angular momentum of spiral galaxies in these models remains a
challenging problem for the $\Lambda$CDM cosmogony.

\section{Summary and Conclusions}\label{sec:sum}

This paper contains the results from 
an extensive suite of numerical simulations which were
aimed at understanding the relationship between the power spectrum of
initial density fluctuations and the concentration of virialized dark
matter halos. These simulations demonstrate that dark halo
concentration depends both on the amplitude of mass fluctuations as
well as on the shape of the power spectrum. A simple model that takes
this into account by defining an effective amplitude as
$\sigma(M)$ times the logarithmic derivative of $\sigma(M)$ with
respect to mass on scales similar to the characteristic mass of the
halo (\ie that enclosed within the radius where the circular velocity
peaks, $r_{\rm max}=2.17 \,r_s$) has been developed. This model
reproduces the mass and redshift dependence of the concentration in
all 7 cosmogonies investigated here, as well as in the 8 different
cosmogonies probed by NFW. It also extends the earlier models of
NFW and Bullock \etal (2000) to power spectra very different from CDM,
including truncated power spectra such as those appropriate for WDM.

These findings are applied to the Milky Way, where observational limits
on the dark mater content within the solar circle can be turned into
constraints on the shape and normalization of the power spectrum. For
the popular $\Lambda$CDM spectrum, the Milky Way halo mass and the
normalization of the power spectrum must satisfy the condition,
$\sigma_8 \, v_{220} \, m_{\rm dark} < 0.8$, where $v_{220}$ is the 
circular velocity
of the halo ($V_{200}$) in units of $220 \, \kms$ and $m_{\rm dark}$
is the upper limit on the mass of dark matter within $8.5$ kpc of the
middle of the Milky Way in units of $4.3 \times 10^{10} M_\odot$. 
For $\sigma_8=0.9$,
the normalization favored from the abundance of galaxy clusters and 
CMB studies, this implies that the Milky Way halo has a circular
velocity significantly smaller than the rotation speed at the solar
circle, $V_{200} < 195 \, \kms$. This finding may have significant
impact on the luminosity function expected in this model, since $195
\, \kms$ halos are much more abundant than their $220 \, \kms$
counterparts (see, e.g., Cole \etal 1994, 2000). 
Gasdynamical simulations including star formation and feedback also
show that, because of their lower concentration relative to the
NS00a,b study, $\Lambda$CDM halos are also roughly consistent with the
zero-point of the I-band Tully-Fisher relation. The slope and scatter
of this relation is also in good agreement with observed values. 

Halo concentrations
in $\Lambda$CDM simulations are much lower than found by NS00a,b, who
had argued that $\Lambda$CDM halos were too concentrated to be
consistent with observations of the dynamics of spiral galaxies. A
reanalysis of their dataset reveals an inconsistency in the
normalization of the power spectrum used in that work. Instead of the
intended $\sigma_8=1.14$, their simulations had an effective
normalization of $\sigma_8\approx 1.6$.  Once this correction is taken
into account both studies yield consistent results.

The set of simulations reported here thus identify and illustrate the
tight relation between power spectrum and halo concentrations. The
application of these results to the Milky Way and I-band Tully-Fisher
relation lifts previous concerns and suggests that the concentration of
$\sigma_8=0.9$ $\Lambda$CDM halos is not clearly incompatible with
observations.

\section*{ACKNOWLEDGMENTS}

We thank Carlos Frenk for helpful discussions.  This work has been
supported by the National Aeronautics and Space Administration under
NASA grant NAG 5-7151, NSF grant 9870151 and by NSERC Research Grant
203263-98. MS and JFN are supported in part by fellowships from the
Alfred P.~Sloan Foundation. MS is also supported by a fellowship from
the David and Lucile Packard Foundation.  This research was supported
in part by the National Science Foundation under Grant
No. PHY94-07194.

\vfil\eject


\begin{thebibliography}{}

\bibitem{aretal} Avila-Reese V., Colin P., Valenzuela O., D'Onghia E.,
Firmani C., 2000, ApJ, submitted (astro-ph/0010525)

\bibitem{bbks} Bardeen J.M., Bond J.R., Kaiser N., Szalay A.S., 1986,
ApJ, 304, 15

\bibitem{bdj} Bell E.F., de Jong R.S., 2001, ApJ, accepted (astro-ph/0011493)

\bibitem{bull} Bullock J.S., Kolatt T.S., Sigad Y., Somerville R.S.,
Kravtsov A.V., Klypin A.A., Primack J.R., Dekel A., 2000, MNRAS,
accepted (astro-ph/9908159)

\bibitem{cdz} Carollo C.M., de Zeeuw P.T., van der Marel R.P.,
Danziger I.J., Qian E.E., 1995, ApJ, 441, L25

\bibitem{semian} Cole S., Aragon-Salamanca A., Frenk C.S., Navarro
J.F., Zepf S.E., 1994, MNRAS, 271, 781

\bibitem{cole2} Cole S., Lacey C., Baugh C.M., Frenk C.S., 2000,
MNRAS, 319, 168

\bibitem{ap3m} Couchman H.M.P., 1991, ApJ, 368, L23

\bibitem{crz} Cretton N., Rix H.-W., de Zeeuw P.T., 2000, ApJ, 536, 319

\bibitem{db} Dehnen W., Binney J.J., 1998, MNRAS, 294, 429

\bibitem{ecf} Eke V.R., Cole S., Frenk C.S., 1996, MNRAS, 282, 263

\bibitem{enf} Eke V.R., Navarro J.F., Frenk C.S., 1998, ApJ, 503, 569

\bibitem{fpbf} Flores R.A., Primack J.R., Blumenthal G.R., Faber S.M.,
1993, ApJ, 412, 443

\bibitem{fp} Flores R.A., Primack J.R., 1994, ApJ, 427, L1

\bibitem{fwde} Frenk C.S., White S.D.M., Davis M., Efstathiou G.,
1988, ApJ, 327, 507

\bibitem{ortg} Gerhard O., 2000, to appear in Galaxy Disks and Disk
Galaxies, ASP Conference Series, J.G. Funes, S.J., and E.M. Corsini,
eds  (astro-ph/0010539)

\bibitem{gjsb} Gerhard O., Jeske G., Saglia R.P., Bender R., 1998,
MNRAS, 295, 197

\bibitem{hd} Hogan C.J., Dalcanton J.J., 2000, Phys. Rev. D, submitted (astro-ph/0002330)

\bibitem{fuzz} Hu W., Barkana R., Gruzinov A., 2000, Phys. Rev. Lett.,
submitted (astro-ph/0003365)

\bibitem{hjs} Huss A., Jain B., Steinmetz M., 1999, MNRAS, 308, 1011

\bibitem{kkbp2} Klypin A.A., Kravtsov A.V., Bullock J.S., Primack J.R.,
2000, ApJ, submitted (astro-ph/0006343)

\bibitem{kkvp} Klypin A.A., Kravtsov A.V., Valenzuela O., Prada F.,
1999, ApJ, 522, 82

\bibitem{kkbp} Kravtsov A.V., Klypin A.A., Bullock J.S., Primack J.R.,
1998, ApJ, 502, 48

\bibitem{ksgb} Kronawitter A., Saglia R.P., Gerhard O., Bender R.,
2000, A\&AS, 144, 53

\bibitem{lc93} Lacey C., Cole S., 1993, MNRAS, 262, 627

\bibitem{lc94} Lacey C., Cole S., 1994, MNRAS, 271, 676

\bibitem{llvw} Liddle A.R., Lyth D.H., Viana P.T.P., White M., 1996,
MNRAS, 282, 281

\bibitem{mcdb} McGaugh S.S., de Blok W.J.G., 1998, ApJ, 499, 41

\bibitem{m} Moore B., 1994, Nature, 370, 629

\bibitem{metal98} Moore B., Governato F., Quinn T., Stadel J., Lake G.,
1998, ApJ, 499, L5

\bibitem{metal99a} Moore B., Ghigna S., Governato F., Lake G., Quinn
T., Stadel J., Tozzi P., 1999a, ApJ, 524, L19

\bibitem{metal99b} Moore B., Quinn T., Governato F., Stadel J., Lake G.,
1999b, MNRAS, 310, 1147

\bibitem{nw} Navarro J.F., White S.D.M., 1993, MNRAS, 265, 271

\bibitem{nfw96} Navarro J.F., Frenk C.S., White S.D.M., 1996, ApJ, 462, 563

\bibitem{nfw97} Navarro J.F., Frenk C.S., White S.D.M., 1997, ApJ,
490, 493

\bibitem{ns97} Navarro J.F., Steinmetz M., 1997, ApJ, 478, 13

\bibitem{nsa} Navarro J.F., Steinmetz M., 2000, ApJ, 528, 607 (NS00a)

\bibitem{nsb} Navarro J.F., Steinmetz M., 2000, ApJ, 538, 477 (NS00b)

\bibitem{peeb} Peebles P.J.E., 1980, The Large Scale Structure of the
Universe. Princeton Univ. Press, Princeton, NJ

\bibitem{fluid} Peebles P.J.E., 2000, ApJ, 534, L127

\bibitem{ps} Press W.H., Schechter P., 1974, ApJ, 187, 425

\bibitem{hwr} Rix H.-W., de Zeeuw P.T., Cretton N., van der Marel
R.P., Carollo C.M., 1997, ApJ, 488, 702

\bibitem{wdm} Sommer-Larsen J., Dolgov A., 2000, ApJ, submitted
(astro-ph/9912166)

\bibitem{ss} Spergel D.N., Steinhardt P.J., 2000, Phys. Rev. Lett., 84, 3760

\bibitem{grapesph} Steinmetz M., 1996, MNRAS, 278, 1005

\bibitem{sn} Steinmetz M., Navarro J.F., 1999, ApJ, 513, 555

\bibitem{sgb} Stompor R., Gorski K.M., Banday A.J., 1995, MNRAS, 277, 1225

\bibitem{grape} Sugimoto D., Chikada Y., Makino J., Ito T., Ebisuzaki
T., Umemura M., 1990, Nat, 345, 33

\bibitem{sugi} Sugiyama N., 1995, ApJS, 100, 281

\bibitem{robs} Swaters R.A., 1999, Ph.D. thesis, Rijksuniversiteit Groningen

\bibitem{smt} Swaters R.A., Madore B.F., Trewhella M., 2000, ApJ, 531, L107

\bibitem{tf} Tully R.B., Fisher J.R., 1977, A\&A, 54, 661

\bibitem{bbn} Tytler D., O'Meara J.M., Suzuki N., Lubin D., 2000,
Physica Scripta, 85, 12

\bibitem{vdbea} van den Bosch F.C., Robertson B.E., Dalcanton J.J., de
Blok W.J.G., 2000, AJ, 119, 1579

\bibitem{vdbs} van den Bosch F.C., Swaters R.A., 2000, AJ, submitted,
(astro-ph/0006048)

\end{thebibliography}
\end{document}